\documentclass[12pt]{aastex}
\usepackage{emulateapj5}
\usepackage{apjfonts}
\usepackage{psfig}
\usepackage{amsmath}
\usepackage{amssymb}
\usepackage{epsfig}

\shorttitle{Finding White Dwarfs with Transit Searches}
\shortauthors{Farmer \& Agol}

\begin{document}


\title{Finding White Dwarfs with Transit Searches}

\author{Alison J. Farmer\altaffilmark{1} and Eric Agol\altaffilmark{1,2}}

\altaffiltext{1}{California Institute of Technology, Mail Code 130-33, 
Pasadena, CA 91125 USA; ajf,agol@tapir.caltech.edu}
\altaffiltext{2}{Chandra Fellow}



\begin{abstract}
We make predictions for the rate of discovery of eclipsing white dwarf--main
sequence (WD--MS) binaries in terrestrial-planet transit searches, 
taking the planned {\em Kepler} and {\em Eddington} missions as examples.
We use a population synthesis model to characterize the Galactic 
WD--MS population, and we find that, despite increased noise due to
stellar variability compared with the typical planetary case, discovery of
$\ga 10^2$ non-accreting, eclipsing WD--MS systems is likely using 
{\em Kepler} and {\em Eddington}, with periods of $2-20$ days and transit 
amplitudes of $\vert \Delta m\vert \sim 10^{-(4\pm0.5)}$ magnitudes.  
Follow-up observations of these systems could accurately test the theoretical
white dwarf mass--radius relation or theories of binary star evolution.
\end{abstract}



\section{Introduction}

As white dwarf stars are comparable in size to the Earth, searches 
for terrestrial planets via transit of their parent stars should be 
capable of discovering white dwarf (WD) binaries as
well. These can be distinguished from terrestrial planet systems
because the WD will induce detectable Doppler shifts in
the spectral lines of a main sequence (MS) companion, whereas an
Earth-like planet will not, due to its smaller mass. 

An interesting coincidence is that the radius of a 
white dwarf in a tight binary is comparable to its Einstein 
radius \citep{mar01}, which means that both gravitational lensing
and eclipse may be important during transit.
When combined with other information about 
the binary, observation of the change of magnitude during transit
may allow measurement of the mass and radius of the 
white dwarf, allowing a test of the relation originally predicted by 
\citet{cha35}.  Recent measurements of white dwarf masses
and radii using atmosphere models \citep{pro98} have modeling uncertainties,
which would be useful to test with another technique,
and the range of WD masses available in binaries is larger than for
single stars.
The four known eclipsing WD--M-dwarf binaries are a good start \citep{mar00}, but a
larger, alternatively selected sample could allow measurement of the
variation of the relation with WD mass, age, or composition.

This paper addresses the question of how many white dwarf--main
sequence (WD--MS) binaries might be found in searches for transiting planets,
with application to the {\em Kepler}\footnote{See http://www.kepler.arc.nasa.gov/} \citep{bor97}
and {\em Eddington}\footnote{See http://astro.esa.int/SA-general/Projects/Eddington/} \citep{hor02} missions.  We concern ourselves
with white dwarfs in detached binaries, since accreting white dwarfs 
can be found by other means. In $\S$ 2 we
discuss the various modes for discovery of white dwarfs and the
surveys that might detect them, then in
$\S$ 3 we describe the expected properties of WD--MS
binaries in our Galaxy.  In $\S$ 4 we 
estimate the number of binaries that might be found in the surveys
mentioned above, in $\S$ 5 we discuss what one might do with any binaries
discovered via this method, and in $\S$ 6 we conclude.

During the preparation of this paper, \citet{sah03} published a
study of near-field WD microlensing for the {\em Kepler}
mission.  In their paper, numerical lightcurves for fiducial systems are
presented, similar to the analytic lightcurves presented in
\citet{ago02}; however, they did not attempt to accurately estimate
the number of detectable systems, the goal of this paper.   

\section{White dwarf binary variations}
\label{sec:wdvar}

First we consider the case in which the WD passes in front of the MS
star (primary transit).  The Einstein radius is $R_{\rm{E}}=[4R_G a]^{1/2}$, 
where $R_G=GM_{\rm{WD}}/c^2$ 
is the gravitational radius for a lens of mass $M_{\rm{WD}}$ and $a$ is 
the semi-major axis of the binary.  A white dwarf in a binary system
has a size, $R_{\rm{WD}}$, which is comparable to the Einstein radius
\begin{equation}
\frac{R_{\rm{WD}}}{R_{\rm{E}}} \simeq 0.7 \left(\frac{M_{\rm{WD}}}{M_\odot}\right)^{-1/2} \left(\frac{a}{0.1{\rm AU}}\right)^{-1/2} 
\left(\frac{R_{\rm{WD}}}{0.01 R_\odot}\right).
\end{equation}
Thus microlensed images (which will appear a distance $\sim R_{\rm{E}}$ from the
center of the WD) may be occulted by the WD, so that we may
have either dimming (favored if  $R_{\rm{WD}} \gg R_{\rm{E}}$, i.e. small
$M_{\rm{WD}}$, $a$) or amplification (if $R_{\rm{WD}} \ll R_{\rm{E}}$,
i.e. large $M_{\rm{WD}}$, $a$).

If the occulting body (here the WD) is much smaller than the occulted body 
(MS star), the microlensing plus occultation equations take on an
exceptionally simple form \citep{ago03}, with the dimming or amplification 
dependent only on the surface brightness immediately behind the WD, so long as
the WD is away from the edge of the MS star. The fractional change in flux 
during a transit is then given by
\begin{equation}
\Delta f_1 \equiv \left(\frac{2R_{\rm{E}}^2-R_{\rm{WD}}^2}{R_{\rm{MS}}^2}\right)
\left(\frac{F_{\rm{MS}}}{F_{\rm{MS}}+F_{\rm{WD}}}\right)\frac{I(r)}{\langle I
  \rangle} \Theta(R_{\rm{MS}}-r),
\label{eq:deltaf}
\end{equation}
where $F_{\rm{MS,WD}}$ are the MS and WD fluxes respectively,
$R_{\rm{MS}}$ is the radius of the MS star, $r$
is the projected distance of the WD from the center of the MS star, $I(r)/\langle I\rangle$ is
the limb-darkened intensity profile of the source normalized to the flux-weighted mean
intensity, 
and $\Theta$ is the step function.

For close ($a \sim 0.1$ AU) WD--MS systems, $|\Delta f_1| \sim 10^{-4}$,
since $R_{\rm{WD}} \sim R_{\rm{E}} \sim 10^{-2} R_{\rm{MS}}$, similar to the transit
depth of a terrestrial planet, though the effect will be a flux increase
when $R_{\rm{WD}}/R_{\rm{E}} < 2^{-1/2}$. In a system of random
inclination $i$ on the sky, the probability of transits along our line
of sight is $\simeq R_{\rm{MS}}/a$, and for a small transiting body, the
transit duration will be $T_{\rm{tr}} \simeq 2 R_{\rm{MS}} \sin \theta /
v_{\rm{orb}}$, where $R_{\rm{MS}} \cos \theta = a \cos i$ and $v_{\rm{orb}}$ is
the relative orbital velocity.

The fractional change of flux when the MS star passes in front of the
WD (secondary transit) is simply given by $\Delta f_2 \equiv
-F_{\rm{WD}}/(F_{\rm{WD}}+F_{\rm{MS}}) < 0$. The transit will obviously be
deeper for younger, more luminous WDs. The luminosity of a typical WD 1--10 Gyr
after birth is of the order $10^{-3}-10^{-5}$ L$_{\odot}$, so that terrestrial planet searches
should be well suited to detecting these events too. Note that these
will be flat-bottomed transits, but will have the same durations and transit probabilities
as the complementary primary transit. If the orbit is
circular, as will be the case for close systems, a given lightcurve will exhibit 
either both primary and secondary transits (though the transit depths
$\Delta f_1$ and $\Delta f_2$ may
be very different) or no transits at all.

There are other types of variations in WD--MS lightcurves, including fluctuations 
as a result of tidal effects, both directly, due to
tidal distortion of the MS star, and indirectly due to increased MS
rotation rate and hence increased stellar variability, which is
described in detail in \S~\ref{sec:var}. In addition, we may see variations
due to irradiation of the MS star by a hot WD in the very youngest WD systems, 
neglected here, or due to flickering if the WD is accreting
at a low level from the MS stellar wind, also discussed in \S~\ref{sec:var}. 

Since the predicted transit amplitudes are of the same
order of magnitude as those for terrestrial planet transits ($\Delta f \simeq |\Delta m|
\sim 10^{-4}$ magnitudes), surveys
designed to search for other earths ought to pick up WD--MS
systems too. To illustrate this, we use as examples the {\em Kepler}
\citep{bor97} and {\em Eddington} \citep{hor02} missions. These satellites will continuously
monitor the brightness of many stars at high photometric precision to
search for periodic dimming characteristic of a transiting planet.

{\em Kepler} will monitor $\sim10^5$ dwarf stars with $9<V<14$ 
in a $10^\circ \times 10^\circ$ field centered on Galactic coordinates
$(l,b)=(69.6^\circ, 5.7^\circ)$, for at least four years. The proposed
{\em Eddington} design is for a smaller $3^\circ \times 3^\circ$ field, with
deeper magnitude limits ($11<V<18$) and a lifetime of 3 years. In the absence of published
coordinates for the {\em Eddington} field, we use here the same $(l,b)$
as {\em Kepler}. {\em Kepler} will have a broad bandpass, extending from $\sim$
400 to 850 nm, while {\em Eddington} may have two-color information \citep{bor03}.
{\em Kepler} will read out fluxes every 15 minutes, and will have a fiducial
sensitivity (similar to that of {\em Eddington}) of $2 \times
10^{-5}$ for a 6.5-hour exposure of a $V=12$ G2V star.

\section{White dwarfs in binaries}
\subsection{Evolution to the WD--MS stage}
\label{sec:evol}

The evolution of a zero-age main sequence (ZAMS) binary system to a white
dwarf--main sequence (WD--MS) system proceeds via one of two main
pathways, according to whether or not Roche lobe overflow
occurs en route, the critical initial orbital separation being at 
$a_{\rm{crit}} \sim 10^3 - 10^4 \rm{R}_\odot$; in both cases, the more massive 
star has evolved into a WD, while the other is still on the MS.
The lifetime of this phase 
depends on the difference in main-sequence lifetimes 
of the two stars.

If the primary fills its Roche lobe on the RGB or AGB, then the
ensuing mass transfer will most likely be dynamically unstable
 and a common envelope phase will
result (see e.g. \citealt{liv93}), in which the envelope of the evolved star is
heated by friction as the secondary and the core of the giant orbit
inside it. This ends when either the stars coalesce or the envelope is heated
sufficiently that it escapes the system, leaving a WD--MS binary with
greatly \emph{reduced} orbital separation compared with the initial ZAMS system. The orbit will be circular, since tidal circularization
will have occurred in any system in which a giant star is close to overflow.

If the orbital separation is sufficiently large that the primary does
not overflow on the RGB or AGB, then it is
affected only as the primary loses its
envelope in the planetary nebula phase, leading to an \emph{increase} in the
orbital separation.  We therefore expect that the Galactic WD--MS population 
will consist of two \emph{distinct} groups of sources in $P-$space, the 
``short'' period systems from systems in which overflow has occurred, and the
``long'' period systems in which no Roche lobe has been filled.

\centerline{\psfig{file=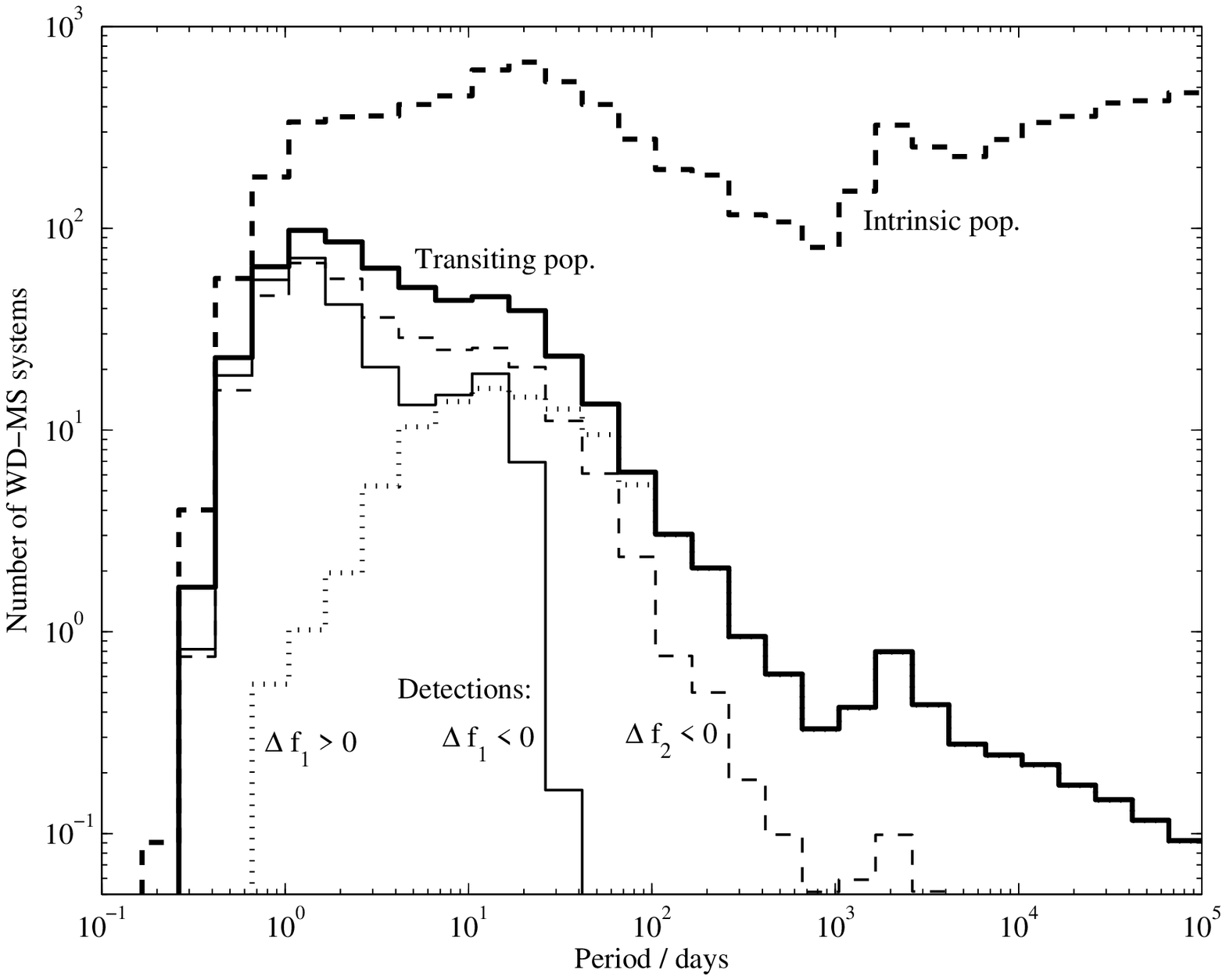,width=\hsize}}

\figcaption{The period distribution of WD--MS systems for {\em Kepler}:
  thick dashed line: intrinsic WD--MS population in the field of view
  (generated as described in \S~3.2); thick
  solid line: intrinsic transiting population along our line of sight; thin lines: detectable
  population using {\em Kepler} (assuming no stellar variability) --- solid
  line: $\Delta f_1 <0$ ; dotted line: $\Delta f_1 >0$, dashed line: $\Delta f_2 <0$.
  Note the dominance of microlensing at longer periods, for which the Einstein radius is
  larger. The introduction of stellar variability (described in
  \S~\ref{sec:var}) limits detectable periods to $\ga 2$d,
  and hampers the detection of shallower transits. \label{fig:period}}


\subsection{Population Synthesis}
\label{sec:assumptions}

A population synthesis approach was used to quantify the properties of
the Galactic WD--MS population. Evolutionary tracks were computed
using the rapid evolution \textsc{bse} code \citep{hur00a}. Following
the preferred Model A of Hurley et al. (2000), we distribute the primary mass according to the
initial mass function (IMF) of \citet{kro93}, while the
secondary mass distribution we choose to be flat in the mass ratio
$q=M_2/M_1$, for $0 < q < 1$. The initial orbital semi-major axis
distribution is flat in $\log a$. These initial conditions, along
with, most notably, a
constant star formation rate over the age of the Galaxy, a binary
fraction of 100\%, solar metallicity across all stars, a common
envelope efficiency parameter\footnote{Note that the common envelope formalism used in Hurley et al. (2000) defines the efficiency parameter $\alpha$ as
$E_{\rm{bind,i}}=\alpha(E_{\rm{orb,f}}-E_{\rm{orb,i}})$, where
$E_{\rm{bind,i}}$ is the binding energy of the envelope, and
$E_{\rm{orb, i, f}}$ are the initial and final orbital energies,
respectively. A smaller $\alpha$ corresponds to a lower ejection 
efficiency and hence a larger loss of orbital energy, and greater 
orbital shrinkage. Here we use $\alpha=3.0$, since this is found by
Hurley et al. (2000) to best reproduce the Galactic double degenerate
population.} $\alpha=3.0$, and zero initial
orbital eccentricity, were found by Hurley et al. (2000) to best reproduce
the observed numbers of double degenerate, symbiotic, cataclysmic
variable and other binary star populations in the Galaxy. Here, we evolve pairs in which both stars have 
masses between 0.1 and 20 M$_{\odot}$, and the 
initial semi-major axis distribution extends from 2 to
$10^5$ ZAMS stellar diameters. Note that the \textsc{bse} code uses the Nauenberg (1972) mass-radius relation for WDs.

We distribute this population according
to a double exponential disk model, with scale length 2.75 kpc. The scale height $h$ is chosen to vary according to stellar age $t$, with $h \propto t^{1/2}$, set equal to 100 pc for stars born today and to 300 pc for the oldest stars in the disk. We use the extinction corrections of \citet{bah80}, and thus model the stars in the {\em Kepler} field of
view within the magnitude limits of the survey. We normalize to the number of dwarf stars counted in this field
($\sim 136,000$ with $9<V<14$, from \emph{Kepler} webpage; results can be re-scaled if this number is later revised).  We estimate that the total number of 
WD--MS systems within this sample of stars is $\sim 15,000$, while the 
{\em Eddington} sample should contain $\sim$ 35,000 WD--MS systems.
The resulting orbital period distribution of target
WD--MS systems is plotted in Figure \ref{fig:period}, and is seen to
display the double-peaked structure expected from \S~\ref{sec:evol}, with a relative dearth of systems with periods from
months to years. 
A similar distribution was predicted by \citet{dek93}.


\centerline{\psfig{file=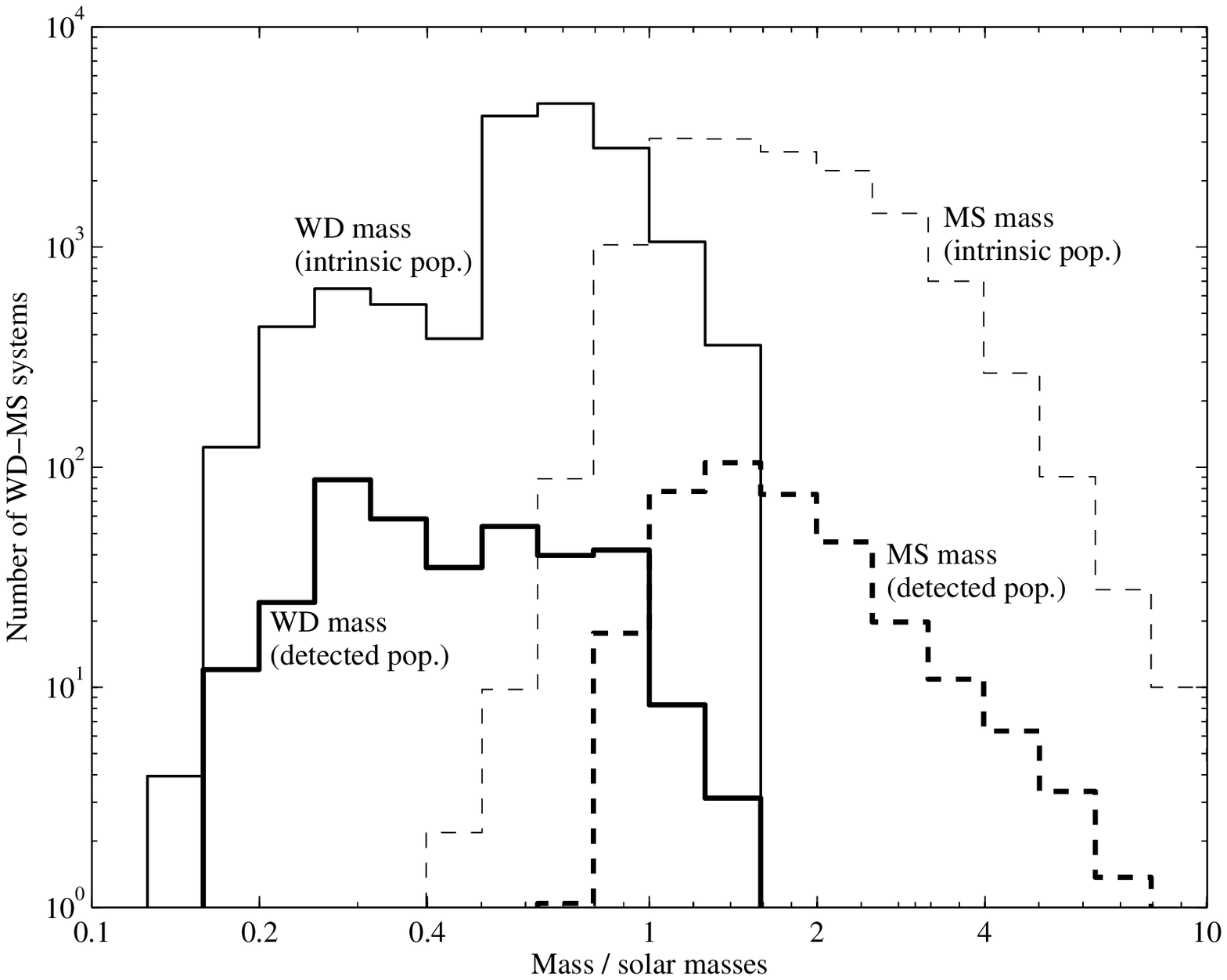,width=\hsize}}

\figcaption{The distribution of MS and WD masses for {\em Kepler}: thin
  solid and dashed lines: intrinsic distributions in WD and MS masses
  respectively,
  for the \emph{Kepler} field WD--MS population. Thick solid and dashed lines: distributions in
  detected systems of WD and MS masses, respectively, where detection
  is of primary transit (WD in front of MS star, $\Delta f_1$ positive or negative). Secondary transit
  distributions are not significantly different. Note that low-mass
  WDs are preferentially detected, both due to their preferential
  formation as short-period systems and their larger physical size,
  making occultations deeper. The MS mass distribution peaks around 1
  M$_\odot$, a region traditionally difficult to survey for WD
  secondaries. \label{fig:mass}}


\section{Detecting white dwarfs}

The detectability depends on the signal-to-noise
ratio of all transits combined in the time series for a given
system. We require that the transit duration be at least 1 hour (four {\it Kepler} time-samples) for detection.

We assume photon counting statistics typically dominate the noise, but
also add in a fractional instrumental noise of $10^{-5}$. For
simplicity, we neglect limb-darkening. 
Shorter period systems are doubly favored, since the time
spent in transit and the probability of a transit in a system with
random orientation both scale as $P^{-2/3}$. For this reason, {\em Kepler}'s
sensitivity --- designed to detect the longer-period
terrestrial systems --- is easily good enough to detect a large fraction
of the transiting WD--MS systems within the magnitude limits at good
signal to noise. The same is not true of \emph{Eddington}, with its deeper
magnitude limits, which restrict most detectable transits to lower-mass
MS primaries. The 
properties of the detectable systems, assuming an $8 \sigma$ detection
threshold, and the parent population from
which they are drawn, are illustrated in Figures \ref{fig:period},
\ref{fig:mass}, and \ref{fig:dm} for {\em Kepler}. The total
numbers of WD--MS systems detectable are summarised for both {\em Kepler} and
{\em Eddington} in Table 1.  We split the transits into the three modes of
detection described in \S~\ref{sec:wdvar}.  Note that many systems
have both detectable primary and secondary transits. For a blind search, a number of transits ($\ga 3$) would need to be seen for detection as a periodic signal amongst the noise, but if radial velocity information were additionally available, it is possible that an identification could be made based on fewer transits. For this reason, systems at all periods are included as detectable (weighted by their transit probabilities and appropriate signal-to-noise ratios); if no such additional information is available, the curves in Figure 1 should be cut off at about $P \sim 1.3$ years, though it can be seen that this makes little difference to the overall number of detectable systems, as a result of the underlying WD--MS period distribution.

We see from Table 1 that several hundred transiting WD--MS
systems are in principle detectable with each mission. However, the most serious 
(and least well-defined) limitation is still to be added:
that of stellar variability noise, which is discussed in the following
section.

\centerline{\psfig{file=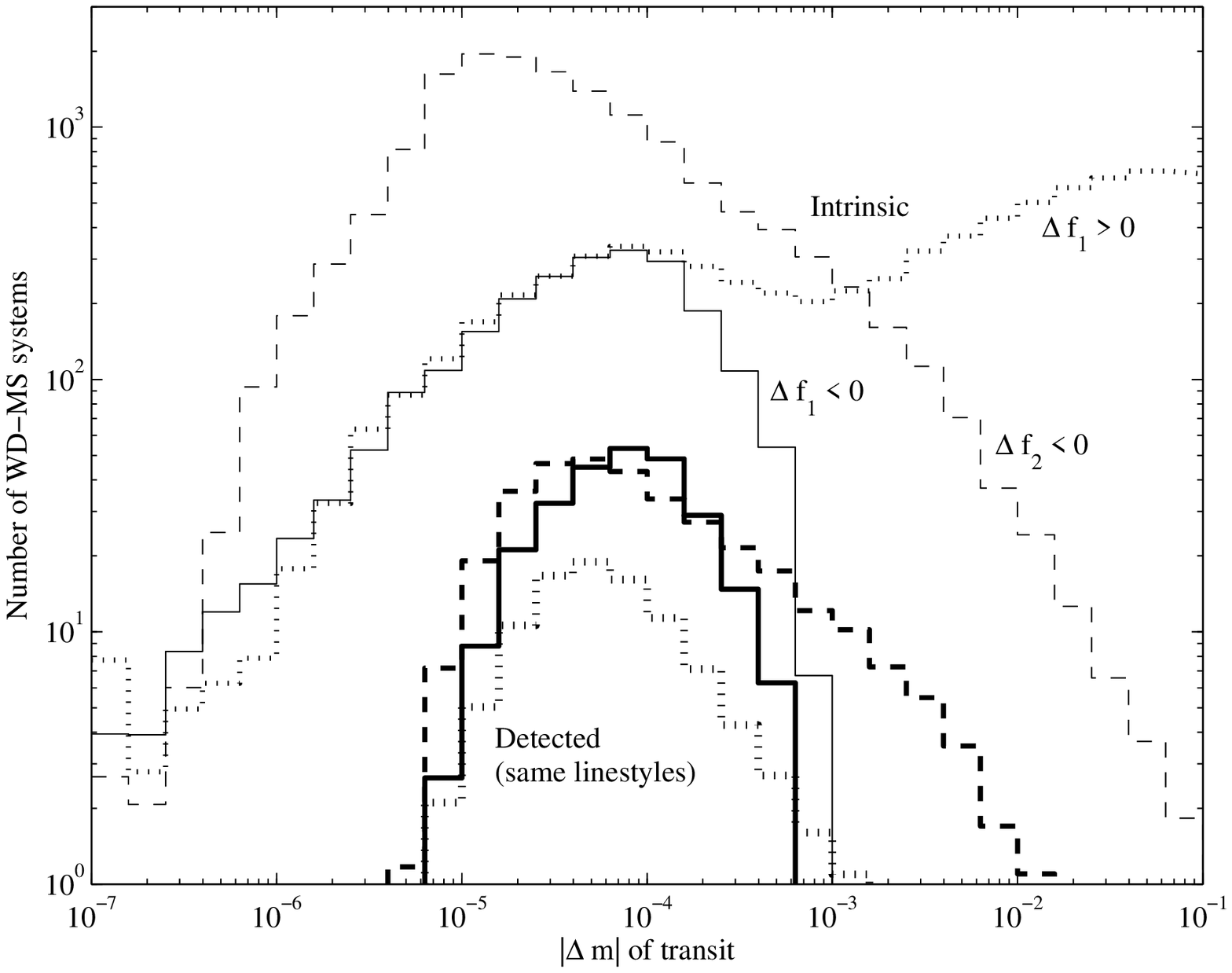,width=\hsize}}

\figcaption{The distribution of transit depths for {\em Kepler}: thin solid,
  dotted and dashed lines show intrinsic WD--MS population displaying
  $\Delta f_1 <0$, $\Delta f_1>0$, and $\Delta f_2<0$, respectively
  (without taking into account transit
  probability along our line of sight). Thick lines show detectable
  population, with the same linestyles, assuming no stellar
  variability. Note the sensitivity cutoff at $\sim 10^{-5}$
  magnitudes' flux variation for all transit types, the tail of
  transits of WD by MS to high $|\Delta m|$ due to young WDs, and the
  non-detectability of the large $|\Delta m|$ values from microlensing
  of MS by WD, since these are long-period systems for which transit
  probabilities and rates are low. \label{fig:dm}}



\subsection{Stellar Variability}
\label{sec:var}

The issue of stellar microvariability is of concern in terrestrial
planet transit surveys (e.g. \citealt{jen02},
\citealt{bat02}, \citealt{aig02}). Variability levels are higher, and
hence more of a problem for WD--MS systems. This is because the
majority of detectable systems (Fig. \ref{fig:period}) have short orbital periods, $P \la 30$d, and at these short periods, tidal effects due
to the WD are significant (note that the \textsc{bse} code follows
tidal effects in detail). Synchronization of the MS star's rotation with the orbital period is
rapid if the MS star has a convective envelope ($M_{\rm{MS}} \la 1.6
\rm{M}_\odot$) and $P \la 10$d. This is the case for about half of the
transits in principle detectable with {\em Kepler} (see Fig. \ref{fig:mass}), and most
for {\em Eddington}. Rapidly
rotating late-type stars display increased starspot activity \citep{mes01}, and hence greater photometric variability, due to these starspots
rotating into and out of sight. Individual spots persist for only a few
rotational cycles, meaning that over time the variability has a random
nature. The amplitude of this variability scales with rotational
period approximately as $\sigma \propto
P_{\rm{rot}}^{-1.5}$. Examination of the power
spectrum of solar irradiance
variations \citep{fro97,jen02} shows that this starspot
noise is present up to frequencies $\sim 25/P_{\rm{rot,}\odot}$.

Of most importance in transit detection is the stellar noise on the
timescale of a transit, since one can in principle filter out variability on other timescales
(e.g. \citealt{jen02}, \citealt{aig02}). For WD--MS systems with $P
\sim 1-10$d, we have $T_{\rm{tr}} \sim 1-3$h. If a WD transit has
$P_{\rm{rot}} \sim P_{\rm{orb}} < 25 T_{\rm{tr}}$, then the fractional starspot noise will be $\gg 10^{-4}$, and will drown
out almost all WD transit signals, even if there are many transits
during the survey lifetime. Effectively this places a lower limit on
the orbital period of detectable systems (of around $P \sim 2$d for
typical systems). At frequencies $\ga 25/P_{\rm{rot}}$, the noise is governed by convective (super-)granulation, has an
amplitude $\sim 5 \times 10^{-5}$ in the Sun on the appropriate
transit timescale, and is likely unaffected by rotation rate. This
noise will however reduce detectability of faint transits.

We approximate all convective stars as solar in these respects, and in
Table 1 show the sizable effects of adding this variability noise upon
the detection rates for {\em Kepler}, again requiring $8 \sigma$ detections. The {\em Eddington} mission, which will
observe in two colors, may be able to use the color signature of the
stellar variability to enhance detection probabilities \citep{bor03},
in which case our predictions without stellar variability may be
more appropriate.

More massive MS stars have radiative envelopes, and smaller 
$\vert \Delta f\vert$ for transits, since their radii are larger. For 
these stars, there is less detailed literature available on stellar
microvariability, so we do not attempt to calculate its expected
impact on the WD--MS detection rate.  We do however note that these
stars can be intrinsically quite variable.
Since the radiative tide is
weaker, systems are most likely asynchronous, but \citet{zaq02} suggest that
in this case, tides can excite the fundamental mode of pulsation of
the star, potentially leading to oscillations on
roughly the timescale of a transit.  We note also that some
observations (e.g. \citealt{dem93}) suggest that stars in close binaries display
greater activity than single stars of the same rotation rate.  Thus,
the numbers of detectable systems given in Table 1 for MS primaries 
with radiative envelopes are likely to be reduced by variability, but we 
have not attempted to quantify this reduction.
Extensive data on these topics may only be 
acquired once space-based transit searches fly.

An additional source of microvariability in the lightcurve may be from flickering as
the WD accretes at a low level from the MS star's wind. To have an
accretion luminosity $L_{\rm{WD}} \sim 10^{-4} \rm{L}_\odot$, the WD
needs to accrete at a rate $\sim 10^{-13} \rm{M}_\odot
\rm{yr}^{-1}$, cf. the solar mass loss rate $\sim 10^{-14} \rm{M}_\odot
\rm{yr}^{-1}$. Only a fraction of the mass lost from the MS star will be accreted
by the WD, though we note that stellar wind mass loss may be enhanced in close
binaries.

\centerline{\psfig{file=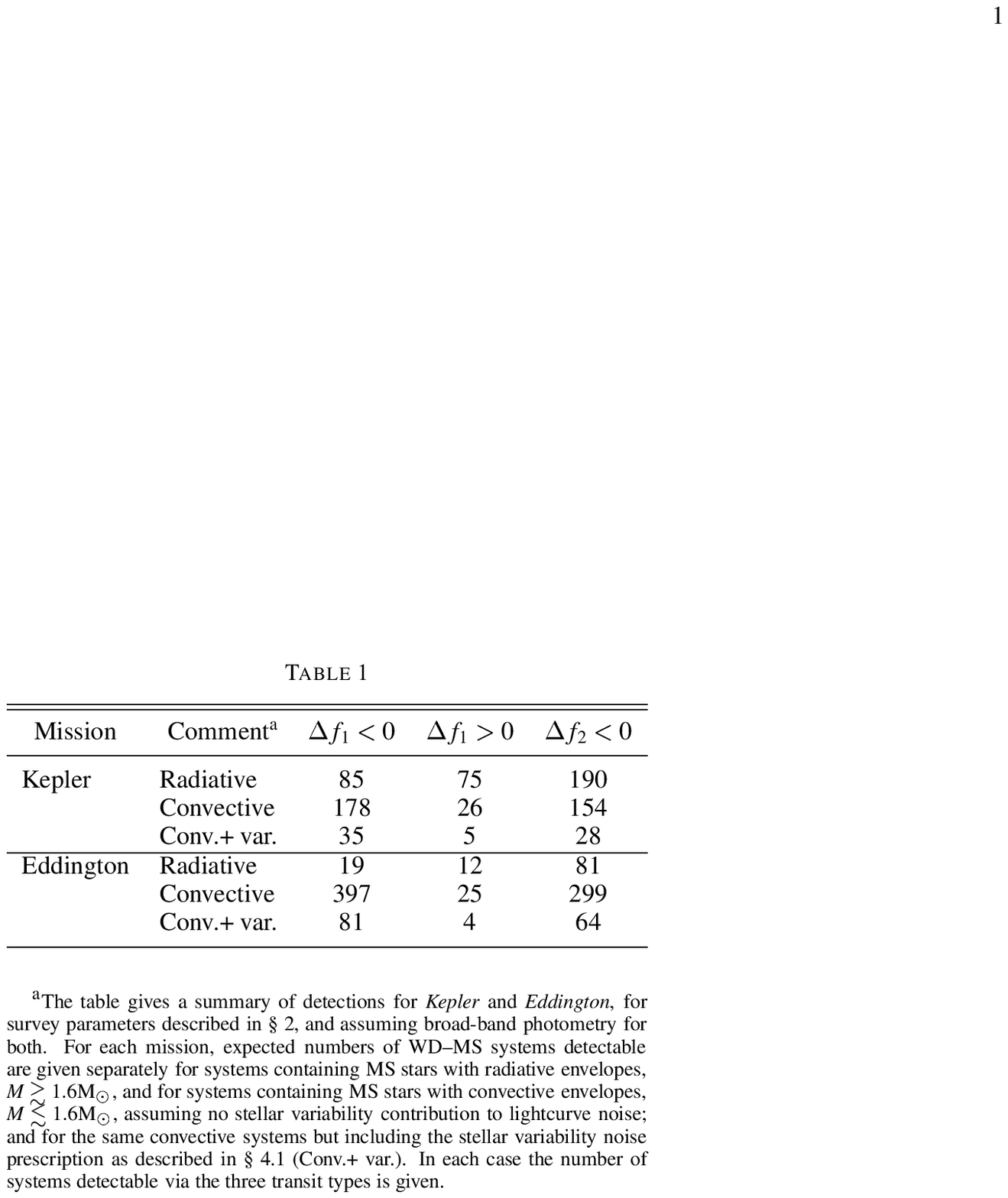,width=\hsize}}

\section{Discussion}

As can be seen from Figure \ref{fig:period}, it is unlikely that
WDs will provide a significant source of spurious earth-like transits
at $P \sim$ years, given the dearth of WD--MS systems and
(hypothesized) large numbers of terrestrial planets at these
periods. In addition, primary transits will be microlensing
events at these periods. Although WD companions are easily
distinguished using radial-velocity observations,
it is useful to know that terrestrial planet signatures will not be
swamped by those of WDs.

A large sample of close WD--MS systems would enable useful tests of
binary star evolution theories (such as common envelope
evolution). Also, if the mass and radius of the WD can be
separately determined, then the WD mass-radius relation could be
tested. The transiting WDs cannot, in
general, be observed other than by their dimming effect upon the MS
star, so that all properties must be inferred. This approach is
however independent of WD atmosphere modeling.

The {\em Kepler} and {\em Eddington} missions easily have the sensitivity to
produce high-quality lightcurves of short-period WD--MS systems, since
they are designed to search for longer-period terrestrial planets. The
inclusion of stellar variability may affect this somewhat. However, there is still a population of
$\ga 50$ WD--MS systems in principle detectable with each, given adequate signal
processing power. Many of these systems display both detectable primary and
secondary transits, which can be distinguished using their transit
profiles. WD transits are longer in duration than, and shaped
differently from, grazing MS--MS
transits of the same depth. Radial velocity measurements should eliminate blending (dilution of a larger transit depth to the expected WD--MS level due to the presence of a brighter star within the same resolution element) as
a source of confusion. This is simpler than in the planetary case, since WDs induce larger radial velocity variations on the orbital timescale. If variability or ellipsoidal
modulation of the MS star flux, radial velocity variations, or characteristic accretion luminosity from the WD, should
draw attention to a system as a candidate close WD--MS system, then
transit searches could also be targeted towards these sources, since
the geometric transit probability can be of order 10\% or
more. Provided none of these is a significant noise source on the
transit timescale, discovery of the transits in the lightcurve is
possible, and system parameters therefore extractable. Although it has been proposed that systems displaying large radial velocity variations be left out of the {\it Kepler} survey (D. Sasselov, private communication, 2003), the findings of this paper strongly
argue for the inclusion of candidate WD--MS systems in the target lists
of transit surveys.

It has been noted by \citet{gou02} that the sensitivity of {\it Kepler} to habitable planets could be significantly increased by pushing the magnitude limit for red MS stars to $V=17$. Such a change would also be expected to increase the survey's sensitivity to WD--MS systems, since the typical orbital periods of transiting systems are shorter, and so the required signal-to-noise is more easily achieved. In addition, we expect there to be a large underlying population of late MS star--WD systems available for detection. However, it should be noted that later-type stars tend to be more photometrically variable, which may complicate the situation.

We need 8 quantities to fully characterize a given system: $M_{\rm{MS}}, R_{\rm{MS}},
M_{\rm{WD}}, R_{\rm{WD}}, P, i, L_{\rm{WD}}$, and $L_{\rm{MS}}$. Orbits
are expected to be circular. The effect of microlensing complicates
the parameter extraction in some sense, since the primary transit depth
depends on both WD mass and radius. We have a number of observables from the
transit lightcurves: $P, T_{\rm{tr}}$, and the primary and secondary
transit magnitude changes, $\Delta m_1$ and $\Delta
m_2$. Ingress/egress times, $\sim$minutes in duration, will be
unmeasurable if the proposed 15 minute exposures of
{\em Kepler} are used. With radial velocity follow-up, we can
also measure $v_{\rm{orb,MS}}$. More information is clearly necessary to
solve for all system parameters. MS modeling can give us
$M_{\rm{MS}}, R_{\rm{MS}}$, and $L_{\rm{MS}}$, at least in
principle, though this may not be accurate for particularly active
stars, or those which have passed through common envelopes. If the
distance to the system is known (for example from {\it Space
  Interferometry Mission}\footnote{See http://sim.jpl.nasa.gov} parallax
measurements) then the situation may be improved. If one is able to
model the limb-darkening of the MS star in the primary transit then
perhaps more information might be extracted. The potentially large
numbers of such systems available in transit surveys make feasible statistical
tests of WD and binary star theory.

\section{Conclusions}

Since the {\em Kepler} and {\em Eddington} missions are designed to
detect terrestrial planets, they are ideal for detecting white dwarfs
as well.   White dwarfs have more modes of detection than
planets due to their larger masses (and luminosities), but their
detection is complicated by the (tidal) effects of these larger masses. In
principle, at least 50 new WD--MS systems might be unambiguously
detected with either mission, while at most 500 could be detected if
stellar variability were less significant than our estimates. 
Follow-up observations of the systems might yield mass and
radius estimates for the (unseen except by transit) WDs, and hence
give a test of the WD mass-radius relation, or of theories of binary
star evolution.

\acknowledgments

We thank J. Hurley for providing the
\textsc{bse} code, and S. Seager for an idea leading to this
work. Support for E. A. was provided by NASA through Chandra Postdoctoral 
Fellowship award PF0-10013. We are grateful to the referee, A. Gould, for helpful and detailed comments.




\begin{thebibliography}{}
\bibitem[Agol(2002)]{ago02} Agol, E., 2002, ApJ, 579, 430
\bibitem[Agol(2003)]{ago03} Agol, E., 2003, ApJ, submitted (astro-ph/0303457)
\bibitem[Aigrain et al.(2002)]{aig02} Aigrain, S., Gilmore, G.,
  Favata, F. \& Carpano, S., 2002, to appear in Conference Proceedings
  "Scientific Frontiers in Research on Extrasolar Planets", ed. Drake
  Dreming (astro-ph/0208529)
\bibitem[Bahcall \& Soneira(1980)]{bah80} Bahcall, J. N. \& Soneira, R. M., 1980, ApJS, 44, 73
\bibitem[Batalha et al.(2002)]{bat02} Batalha, N. M., Jenkins, J.,
  Basri, G. S., Borucki, W. J. \& Koch, D., G., 2002, Proc. 1st
  {\em Eddington} Workshop, eds Favata, F., Roxburgh, I. W. \&
  Galad\'{i}-Enr\'iquez, D.
\bibitem[Bord\'e et al.(2003)]{bor03} Bord\'e P., Leger A., Rouan D.,
  Cameron A. C., 2003, submitted to A\&A, astro-ph/0301430
\bibitem[Borucki et al.(1997)]{bor97} Borucki, W. J., Dunham, E. W.,
  Koch, D. G., Cochran, W. D., Rose, J. A., Cullers, K., Granados,
  A. \& Jenkins, J. M., 1997, ASP Conf Ser, 119, 153, ed. Soderblom, D.
\bibitem[Chandrasekhar(1935)]{cha35} Chandrasekhar, S., 1935, MNRAS, 95, 207
\bibitem[de Kool \& Ritter(1993)]{dek93} de Kool, M., \& Ritter, H.,
  1993, A\&A, 267, 397
\bibitem[Dempsey et al.(1993)]{dem93} Dempsey, R. C., Bopp, B. W.,
  Henry, G. W. \& Hall, D. S., 1993, ApJS, 86, 293
\bibitem[Fr\"{o}hlich et al.(1997)]{fro97} Fr\"{o}hlich, C. et al., 1997,
  SoPh, 170, 1
\bibitem[Gould, Pepper \& DePoy(2002)]{gou02} Gould, A., Pepper, J., DePoy, D., L., 2003, ApJL, submitted (astro-ph/0211547)
\bibitem[Horne(2002)]{hor02}Horne, K., 2002, Proc. 1st {\em Eddington}
  Workshop, eds Favata, F., Roxburgh, I. W., \&
  Galad\'{i}-Enr\'{i}quez, D.
\bibitem[Hurley, Pols \& Tout(2000)]{hur00a} Hurley, J. R., Pols, O. R. \& Tout, C. A., 2000, MNRAS, 315, 543
\bibitem[Jenkins(2002)]{jen02} Jenkins, J. M., 2002, ApJ, 575, 493
\bibitem[Kroupa, Tout \& Gilmore(1993)]{kro93} Kroupa, P., Tout,
  C. A., \& Gilmore, G., 1993, MNRAS, 262, 545
\bibitem[Iben \& Livio(1993)]{liv93} Iben, I., Livio, M., 1993, PASP, 105, 1373
\bibitem[Marsh(2000)]{mar00} Marsh, T. R., 2000, NewAR, 44, 119
\bibitem[Marsh(2001)]{mar01} Marsh, T. R., 2001, MNRAS, 324, 547
\bibitem[Messina, Rodon\`{o} \& Guinan(2001)]{mes01}Messina, S., Rodon\`{o},
  M. \& Guinan, E. F., 2001, A\&A, 366, 215
\bibitem[Nauenberg(1972)]{nau72} Nauenberg, M., 1972, 175, 417
\bibitem[Provencal et al(1998)]{pro98} Provencal, J. L., Shipman, H. L., Hog, E., \& Thejll, P., 1998, ApJ, 494, 759
\bibitem[Sahu \& Gilliland(2003)]{sah03} Sahu, K. \& Gilliland, R. L., 2003,  ApJ, 584, 1042
\bibitem[Zaqarashvili, Javakhishvili \& Belvedere(2002)]{zaq02}
  Zaqarashvili, T., Javakhishvili, G. \& Belvedere, G., 2002, 579, 810
\end{thebibliography}
\end{document}